\documentclass[pdftex,aps,prl,amsmath,amsfonts,amssymb,superscriptaddress,twocolumn,10pt]{revtex4}
\renewcommand{\vec}[1]{\mathbf{#1}}
\usepackage[pdftex]{graphicx}
\usepackage{times}
\usepackage{wasysym}
\usepackage{setspace}
\usepackage{color}

\renewcommand{\Im}{\operatorname{Im}}

\newcommand{\figref}[1]{Fig.~\ref{fig:#1}}
\newcommand{\Figref}[1]{Figure~\ref{fig:#1}}

\renewcommand{\eqref}[1]{Eq.~(\ref{eq:#1})}

\newcommand{\citeasnoun}[1]{Ref.~\onlinecite{#1}}

\def\a{s}
\def\b{s}
\newcommand{\add}[1]{\if\a\b{{\color{red} #1}}\else{#1}\fi}
\newcommand{\comm}[1]{\if\a\b{{\color{blue}\{\small \sc #1\}}}\else{}\fi}
\newcommand{\del}[1]{{\if\a\b{{\color{magenta}[[#1]]}}\else{}\fi}}

\begin{document}

\title{Stable suspension and dispersion-induced transitions \\ from
repulsive Casimir forces between fluid-separated cylinders}

\author{Alejandro~W. Rodriguez}
\affiliation{Department of Physics,
Massachusetts Institute of Technology, Cambridge, MA 02139}
\author{J.~N. Munday}
\affiliation{Department of Physics, Harvard University, Cambridge, MA 02139}
\author{J.~D. Joannopoulos}
\affiliation{Department of Physics,
Massachusetts Institute of Technology, Cambridge, MA 02139}
\author{Federico Capasso}
\affiliation{School of Engineering and Applied Sciences, Harvard University, Cambridge, MA 02139}
\author{Diego A.~R. Dalvit}
\affiliation{Theoretical Division, Los Alamos National Laboratory, Los Alamos, NM 87545}
\author{Steven G. Johnson}
\affiliation{Department of Mathematics,
Massachusetts Institute of Technology, Cambridge, MA 02139}

\begin{abstract}
  We numerically demonstrate a stable mechanical suspension of a
  silica cylinder within a metallic cylinder separated by ethanol, via
  a repulsive Casimir force between the silica and the metal.  We
  investigate cylinders with both circular and square cross sections,
  and show that the latter exhibit a stable orientation as well as a
  stable position, via a method to compute Casimir torques for finite
  objects.  Furthermore, the stable orientation of the square cylinder
  undergoes a $45^\circ$ transition as the separation length-scale is
  varied, which is explained as a consequence of material dispersion.
\end{abstract}

\maketitle

\begin{figure}[t]
\includegraphics[width=0.97\columnwidth]{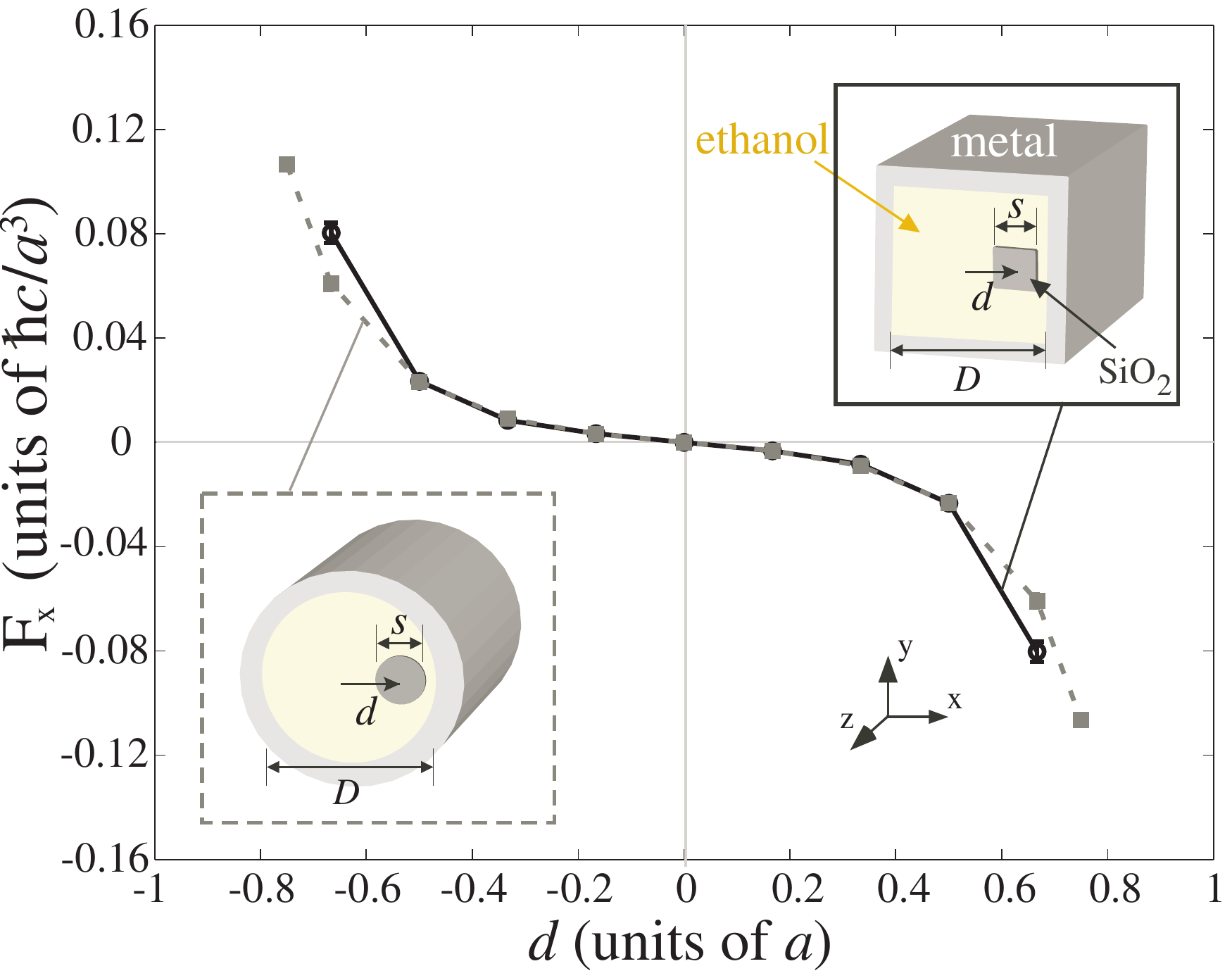}
\caption{Casimir force $F_x$ in the $x$-direction, per unit
$z$-length, on a SiO$_2$ cylinder suspended within a perfectly-metallic
cylinder (inset), separated by fluid (ethanol), as a function of the
$x$-displacement from equilibrium (eccentricity) $d$ [in units
of $a=0.5(D-s)$] for both circular (solid-line) and square
(dashed-line) cylinders. $d=0$ is seen to be a stable equilibrium.}
\label{fig:2d-rep}
\end{figure}

The Casimir force, arising from quantum fluctuations of the
electromagnetic field, was first described as an attractive,
monotonically decaying force between metallic plates~\cite{casimir},
but repulsive interactions can arise in special circumstances,
e.g. involving fluid-separated asymmetric
plates~\cite{Dzyaloshinskii61}. It has been proposed that repulsive
Casimir forces between fluid-separated objects can lead to stable
mechanical equilibria, and hence frictionless static bearings or other
interesting passive-suspension devices~\cite{Munday05,
Capasso07:review}.  However, previous calculations and experiments
involving fluid-separated objects have been restricted to geometries
involving parallel plates or approximations thereof~\cite{Munday07}
(similar to work on air-separated metals, as reviewed in several
recent papers, e.g.~\citeasnoun{bordag01}). Here, using recently
developed numerical techniques~\cite{Rodriguez07:PRA}, we present
theoretical calculations of Casimir forces/torques between
fluid-separated objects with finite square and circular cross sections
(\figref{2d-rep} insets) that rigorously demonstrate stable
positional/orientational Casimir equilibria. (Vacuum-separated
perfect-metal circular cylinders were shown to exhibit an unstable
positional equilibrium~\cite{Dalvit06}.)  In the case of square cross
sections, a surprising result is obtained for the rotational
equilibria: the stable orientation changes by $45^\circ$ depending on
the lengthscale, which we show to be a consequence of material
dispersion. In particular, for certain fluid-separated materials, the
frequency dependence of the permittivity $\varepsilon$ (material
dispersion) causes the Casimir force to switch from repulsive to
attractive at some critical separation~\cite{Dzyaloshinskii61}; this
leads to the orientation transition described here, and may produce
other lengthscale-based qualitative transitions in future
geometries. We present a method to compute Casimir torques on finite
objects (in contrast to previous uncontrolled
approximations~\cite{Jaffe05:approach} or analytical results for
planar cases~\cite{Parsegian72, Enk95:torque, Shao05,
Razmi05, Rodrigues06:torque, Munday05}); we supplement the accurate
results with a heuristic model based on the proximity force
approximation (PFA) that turns out to capture qualitative behaviors of
the orientation transition, and therefore provides some simple
insight.


A repulsive Casimir force arises between parallel plates of
$\varepsilon_{1,2}$ separated by a fluid of
$\varepsilon_f$, if $\varepsilon_1(i\xi) < \varepsilon_f(i\xi) <
\varepsilon_2(i\xi)$ for a sufficiently wide range of imaginary
frequencies $\xi = \Im\omega$~\cite{Dzyaloshinskii61}.  Three such
materials are silica (SiO$_2$) and metal (Au) separated by
ethanol~\cite{Capasso07:review}, as discussed below. We studied
the three-dimensional constant cross-section geometries shown in the
inset of \figref{2d-rep}: square or circular SiO$_2$ cylinders of
diameter $s$ surrounded by a metal cylinder of diameter $D$, separated
by ethanol.  For computational ease, we use a perfect metal
[$\varepsilon(i\xi) \to \infty$] for the outer cylinder, but we use
experimental $\omega$-dependent $\varepsilon$ for the SiO$_2$ and
ethanol. In such cases, with object sizes and curvatures
comparable to their separation, approximations of the Casimir force
as a pairwise attraction between surfaces are not valid and their
qualitative predictions (e.g. stability) may be
incorrect~\cite{Dalvit04, gies06:PFA, Rodriguez07:PRA}.

\Figref{2d-rep} shows the force per unit $z$-length on the inner
cylinder as a function of the displacement $d$ from equilibrium in
units of $a$, where $a \equiv 0.5 (D-s) = 0.0955\,\mu$m so that $d/a =
\pm 1$ for touching surfaces, for parameters $s/D=0.25$. It
demonstrates a stable equilibrium for both square (solid black) and
circular (dashed grey) cross sections.  The computational method is
based on integration of the mean electromagnetic stress tensor (valid
even for fluids~\cite{Pitaevski06}) evaluated in terms of the
imaginary-frequency Green's function via the fluctuation-dissipation
theorem~\cite{Rodriguez07:PRA}. Additional details are provided below.

\begin{figure}[t]
\includegraphics[width=0.97\columnwidth]{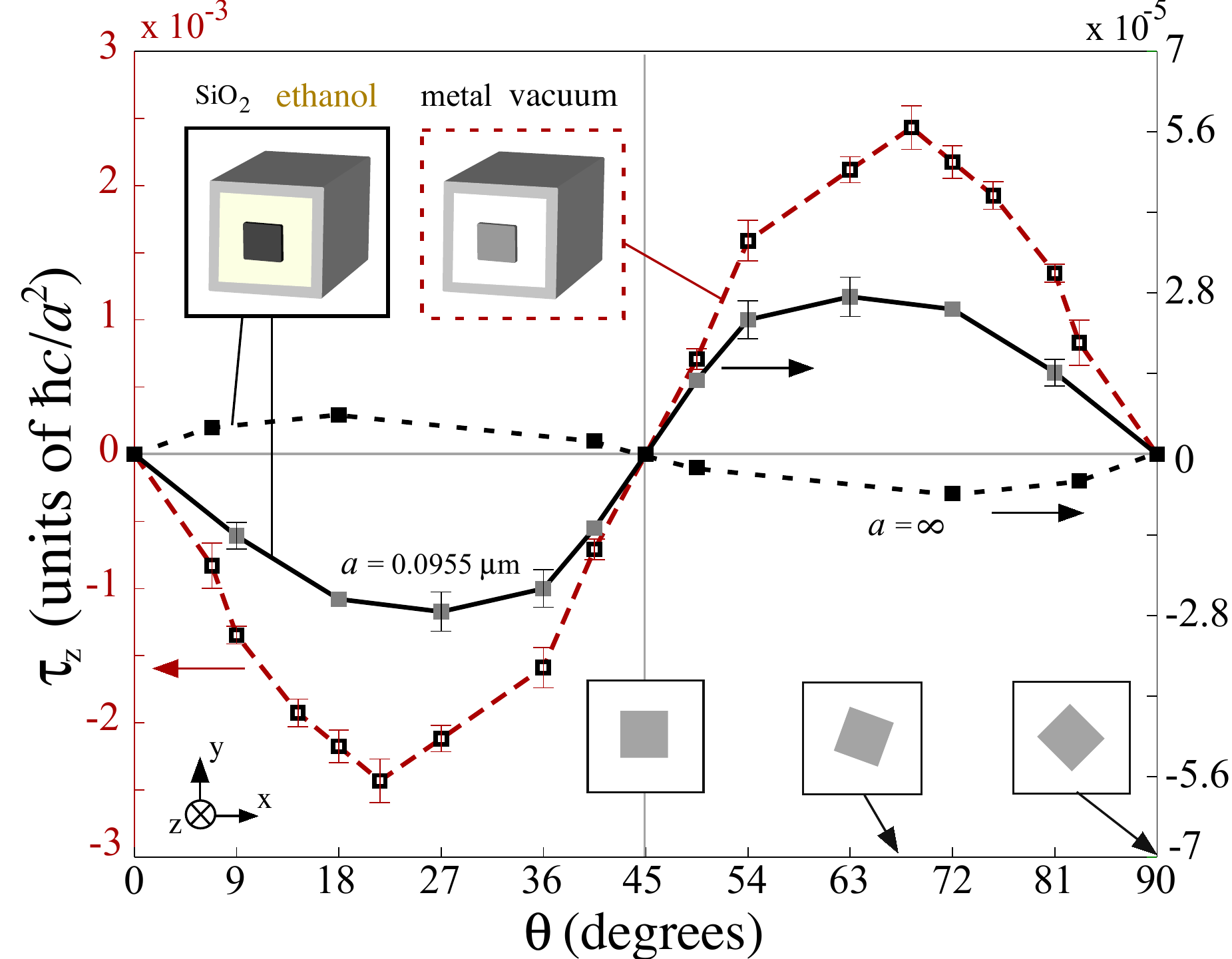}
\caption{Casimir torque $\tau_z$, per unit $z$-length, on the inner
square rod of eccentric square cylinders as a function of the angle
$\theta$ with respect to the $x$-axis (see right insets) for two
material choices as shown in the top insets. Error bars are estimates
of the effect of finite grid resolution.}
\label{fig:torque} 
\end{figure}

Having demonstrated positional stability (\figref{2d-rep}), we now
explore an unusual effect arising from material dispersion.  The
square-cylinder geometry (\figref{torque} inset) must, by symmetry,
exhibit two equilibrium orientations ($\theta=0^\circ$ and
$\theta=45^\circ$). {\it A priori}, it is not clear which of the two
configurations is stable. To determine this, \figref{torque} plots the
Casimir torque per unit $z$-length on the inner square, as a function
of the rotation angle $\theta$ between the inner and outer squares.
In addition to computing the torque for the SiO$_2$--ethanol--perfect-metal case,
we also analyze the torque for vacuum-separated perfect-metal cylinders, in which
case the forces are purely attractive and there is no stable
positional equilibrium~\cite{Dalvit06}.

\begin{figure}[t]
\includegraphics[width=0.97\columnwidth]{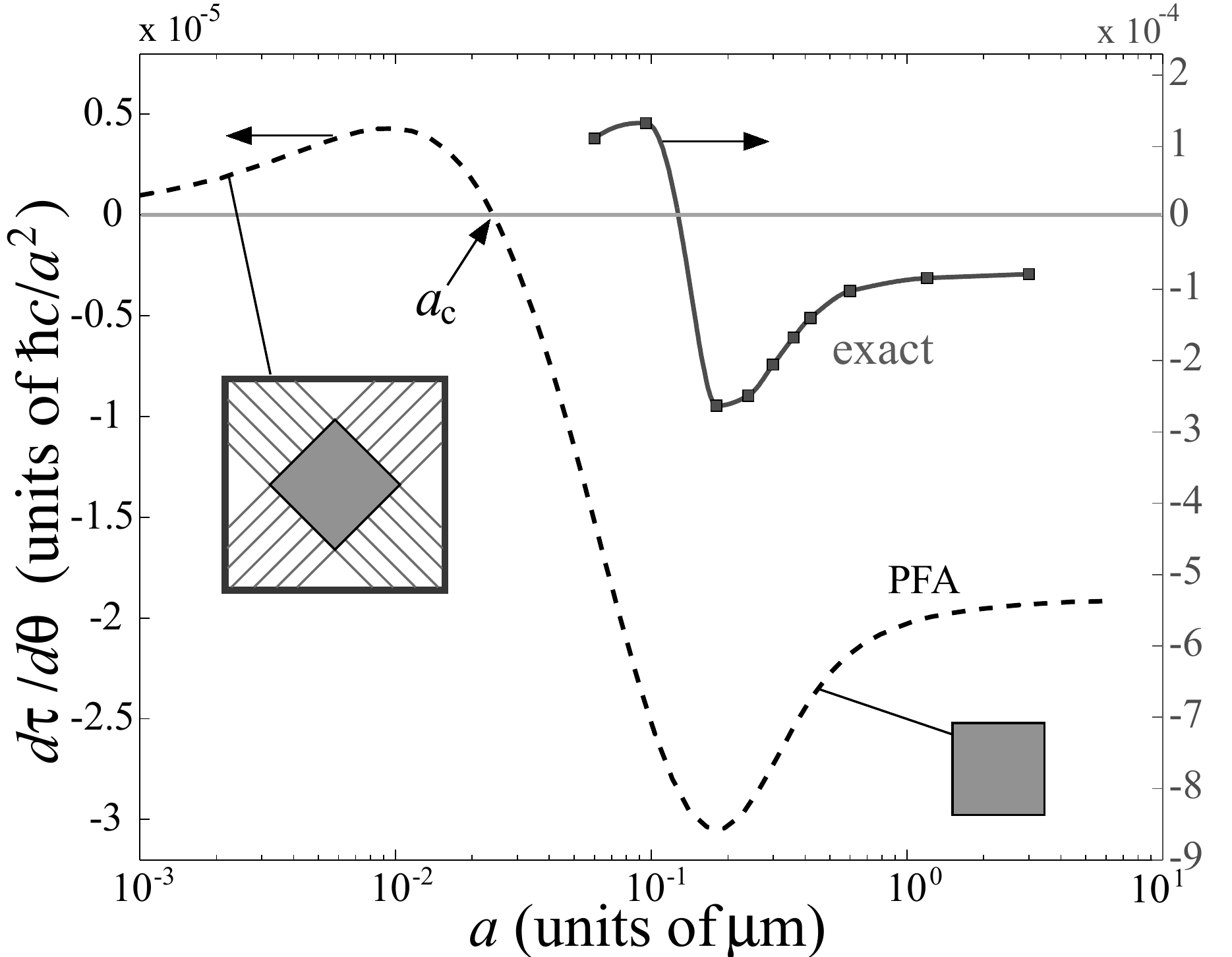}
\caption{Derivative of the Casimir torque in the $z$-direction with
respect to $\theta$, $d\tau_z/ d\theta$, in units of $\hbar c / a^2$,
evaluated at $\theta \approx 0^\circ$, as a function of lengthscale
$a=0.5(D-s)$. The solid line is a fit to the exact Casimir torque,
computed by our numerical method (solid squares), and the dashed line
is the torque as computed by a PFA approximation. Both curves display
a change (dispersion-induced transition) in the stable orientation of
the square.}
\label{fig:transition}
\end{figure}

The resulting torque per unit $z$-length is shown in \figref{torque},
for both the repulsive (solid, filled-square-lines) and attractive
(solid, open-square-line) cases, as $\theta$ is varied from
$-45^{\circ}$ to $45^{\circ}$. The stable orientation for both cases
is $\theta=45^\circ$, a surprising result considering the difference
in the sign of the force. A stable orientation of $45^\circ$ in the
perfect-metal case is not so surprising, since an attractive force
that decreases with separation should intuitively (in the heuristic
picture of pairwise attractions between surfaces) favor an orientation
where the surfaces are as close as possible, pulling the corners of
the inner square towards the outer surfaces.  On the other hand, a
repulsive force should intuitively push the surfaces as far apart as
possible, which would suggest a $0^\circ$ stable orientation.  The
reason for this apparent contradiction between intuition and the
numerical results lies in the effects of material dispersion---as
explained below, $\varepsilon_\mathrm{Si}-\varepsilon_\mathrm{eth}$
(and hence the force) switches sign at some lengthscales (some
$1/\xi$). For example, if we neglect the material dispersion of
SiO$_2$ and ethanol, and use only the $\xi \rightarrow 0$ dielectric
constants, the torque as a function of $\theta$ (shown as the dashed
black line in \figref{torque}) indeed exhibits a stable orientation at
$0^\circ$, as expected from the pairwise intuition: in this case, the
force is repulsive at \emph{all} lengthscales.

Since $\theta=0^\circ$ is unstable for the $a=0.0955\, \mu$m fluid
case and stable for $a \rightarrow \infty$ (equivalent to $\xi
\rightarrow 0$), there \emph{must} be a transition at some critical
intermediate lengthscale $a_c$. One way to determine $a_c$ is to
calculate the derivative of the torque at $\theta=0^\circ$ as a
function of $a$ (noting that the dimensionless torque $\tau a^2 /
\hbar c$ and its derivative go to a nonzero value as $a\to\infty$),
and to look for a change in the sign.  This derivative is plotted as
the solid line in \figref{transition} and displays the expected
transition from unstable ($d\tau_z/d\theta > 0$, left) to stable
($d\tau_z/d\theta < 0$, right) at $a_c \approx 0.1\,\mu$m, a
consequence of material dispersion.  A better understanding of this
transition can be gained by inspecting a simple heuristic (PFA).  PFA
is only an {\it ad hoc} model, in which the force on each point of the
surface (and hence the torque) is treated as simply the parallel-plate
(Lifshitz) force between fluid-separated half-spaces (with ``lines of
interaction'' perpendicular to the inner square, as depicted in the
left inset of \figref{transition}). This simple model turns out to
capture some qualitative features of the orientation transition, as
shown by the dashed line in \figref{transition}, although it is of
course quantitatively incorrect~\cite{Rodrigues06:torque}.

PFA provides an explanation for why orientational stability need not
coincide with positional stability. Stability, in general, arises from
the competing interactions of the inner and outer surfaces, e.g. in
the $45^\circ$ orientation for repulsive interactions, the nearest
surface ``pushes'' the corner away, while the other surface ``pushes''
the corner back. Which of these competing forces dominates depends on
their sign and power law, but in the torque $\boldsymbol{\tau}=\vec{r}
\times \vec{F}$ case, there is an additional effect (from
the $\vec{r} \times {}$ dependence): the force ``from'' the nearest
surface elements is parallel to the radial direction and does \emph{not}
produce a torque, whereas it \emph{does} contribute to the total
force. Therefore, the net force and torque can respond
differently to the same material dispersion, because the lengthscales
of the dominant contributing separations differ.  More generally, in
the exact model where the force cannot be decomposed into additive
contributions, the dominant imaginary-frequency contribution can
differ between torque and force.

\begin{figure}[t]
\includegraphics[width=0.95\columnwidth]{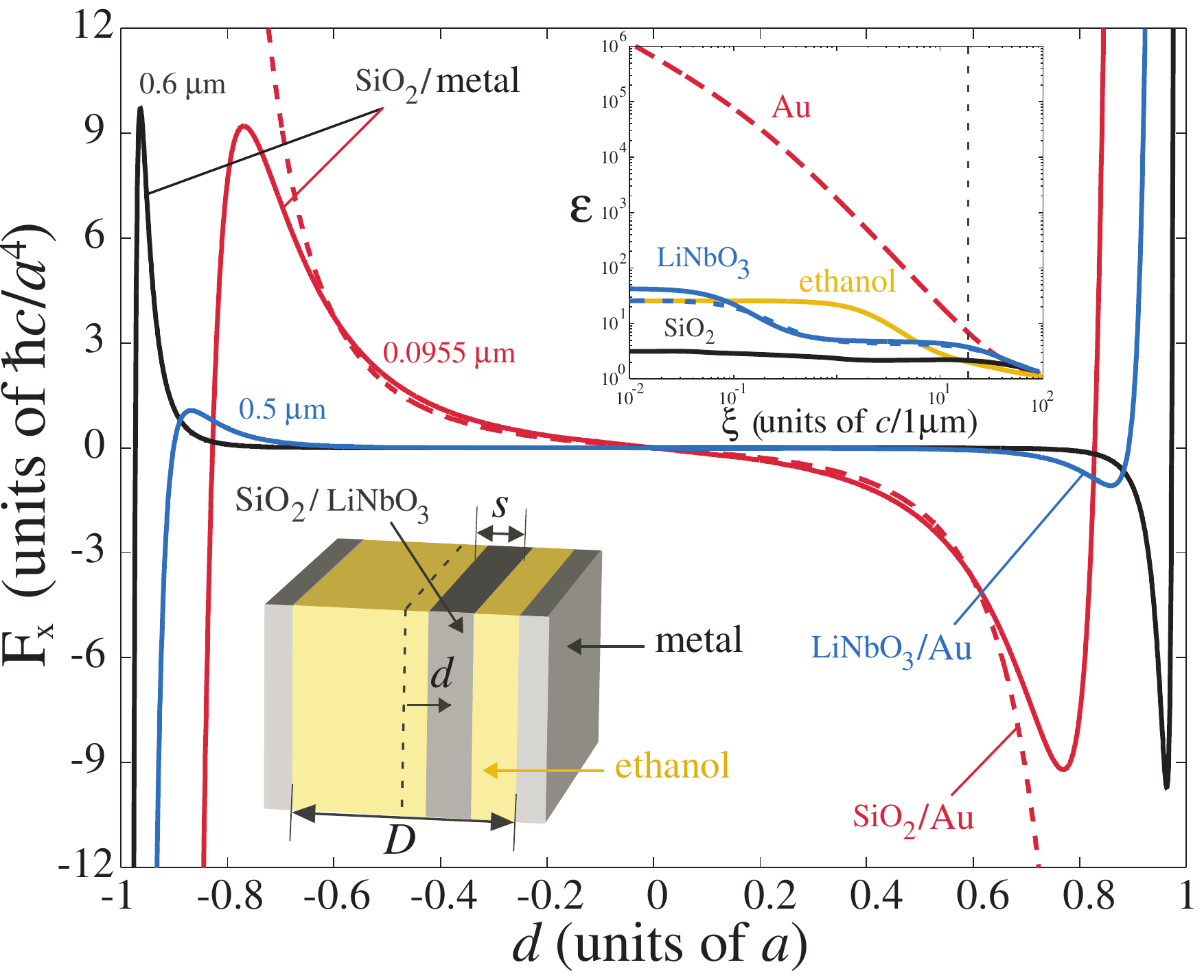}
\caption{Casimir force $F_x$ in the $x$-direction, per unit area,
between a planar SiO$_2$ slab suspended between two perfect-metal
plates (solid red and black lines) or gold half-spaces (dashed red
line), separated by a fluid (ethanol), as a function of the
dimensionless $x$-displacement $d$ from equilibrium. The force is
plotted also at two lengthscales $a \equiv 0.5(D-s)=0.0955\,\mu$m (red
lines) and $a=0.6 \, \mu$m (black line). The solid blue line shows the
same quantity for the case of a lithium-niobate slab and a gold
half-space, for $a=0.5 \, \mu$m. (\emph{Insets:}) Top inset:
$\varepsilon(i\xi)$ for SiO$_2$ (black), LiNbO$_3$ (solid and dashed
blue), ethanol (orange) and gold (red) as a function of imaginary
frequency $\xi$. Bottom inset: schematic of geometry.}
\label{fig:1drep}
\end{figure}

There is another interesting feature in \figref{transition}: for $a >
a_c$, the derivative of the torque is nonmonotonic, \emph{decreasing}
in magnitude towards $a\to\infty$.  This may seem
counter-intuitive because the $\varepsilon$ contrast is maximum for
$\xi\to 0$ (\figref{1drep} inset), which would seem to predict greater
forces.  However, because the net torque arises from a competition
between nearby and faraway surfaces, the magnitude depends not only on
the force but also on the power law: if the force decreases more
rapidly with distance, then faraway surfaces contribute less and the
net torque is larger.  This is precisely what is happening here:
at intermediate $a$, material dispersion acts to increase the
power law~\cite{Dzyaloshinskii61} compared to the $a\to\infty$ case
where dispersion plays no role.  Preliminary work shows similar
effects in the positional stability.  This also explains the relative
torque magnitudes for $a \rightarrow \infty$ and $a=0.0955\, \mu$m in
\figref{torque}.

Since the qualitative behavior of the orientation transition is captured by PFA, it makes
sense to examine a simple one-dimensional model in detail, depicted on
the inset of \figref{1drep}: a SiO$_2$ slab of thickness $s$,
separated from two metallic half-spaces, a (surface-to-surface)
distance $a\equiv 0.5 (D-s)$ from the SiO$_2$ surface, with ethanol
(fluid) in between.  This geometry is analyzed using a generalization
of the Lifshitz formula~\cite{Tomas02}.  \Figref{1drep} shows the
force per unit area on the inner slab (again using $s/D=0.25$), as a
function of the displacement $d$, for two lengthscales: $a=0.0955\,
\mu$m (red), and $a=0.6\, \mu$m (black).  There is a stable
equilibrium at $d=0$, just as for the two-dimensional case in
\figref{2d-rep}.  In the case of perfect-metal half-spaces, the force changes
sign at a critical distance $d_c$ that depends on the lengthscale $a$.
However, if we use a realistic metal (Au), shown by the dashed line
for $a=0.0955\, \mu$m, this transition disappears and the force is
always repulsive.  These features are explained by the inset of
\figref{1drep}, which shows $\varepsilon(i\xi)$ for SiO$_2$, ethanol,
and Au. The key point is that the contributions to the Casimir force
come primarily from imaginary ``wavelengths'' $2\pi/\xi$ larger than
some lengthscale set by the separation, while very short wavelengths
(large $\xi$) on this scale are exponentially cut off in the force
integral~\cite{lifshitz1}. Thus, for large lengthscales $a$,
the force is dominated by small-$\xi$ contributions where
$\varepsilon_\mathrm{silica} < \varepsilon_\mathrm{ethanol} <
\varepsilon_\mathrm{metal}$, giving a repulsive force and a stable
equilibrium.  On the other hand, for small lengthscales and
separations, large-$\xi$ contributions become more important, for
which $\varepsilon_\mathrm{silica} > \varepsilon_\mathrm{ethanol}$
(for $\xi > 2.3\,c/\mu$m, marked by the vertical dashed line in the
inset of \figref{1drep}), leading to attractive forces.  For the case
of perfect-metal cylinders, these attractive contributions are large enough to
flip the sign of the total force for small separations, whereas for Au
the attractive high-$\xi$ contributions are suppressed by the
diminishing $\varepsilon(i\xi)$ of the Au and the force remains
repulsive.  While the transition is absent in the SiO$_2$-ethanol-Au
configuration, we have calculated the force between ethanol-separated
lithium-niobate LiNbO$_3$ and Au plates and found a sign transition at
separations $\sim 0.05 \, \mu$m.  This is illustrated in
\figref{1drep} (solid blue line); because LiNbO$_3$'s anisotropy
greatly complicates the modeling, we used a PFA approximation based on
summing the Lifshitz forces between semi-infinite Au and LiNbO$_3$
slabs (neglecting the finite LiNbO$_3$ thickness).  The two principal
values of the $\varepsilon$ tensor of LiNbO$_3$~\cite{Bergstrom97} are
plotted in the inset.  There are also transitions for
ethanol-separated barium-titanate and calcite plates at separations
$\sim 0.01\,\mu$m~\cite{Munday05}.  Since these two cases are much
more difficult to compute (the outer cylinder is not perfect metal and LiNbO$_3$
is anisotropic), we focused on the SiO$_2$-ethanol-metal case
here, which has similar qualitative behaviors.

The experimental $\varepsilon$s of SiO$_2$ and ethanol were fit to a
standard multiple-oscillator model~\cite{Mahanty76} accurate
from infrared to ultraviolet wavelengths:
$\varepsilon(i\xi) = 1 + \sum_{n=1}^N C_n
\left[1+\left(\xi/\omega_n\right)^2\right]^{-1}$.  The parameters we
used are~\cite{Milling96, Bergstrom97}: ethanol ($N=2$) $\omega_n = \{
6.6, 114 \} \times 10^{14}$~Hz and $C_n = \{23.84,0.852\}$; SiO$_2$
($N=3$) $\omega_n = \{ 0.867, 1.508, 203.4 \} \times 10^{14}$~Hz and
$C_n = \{ 0.829, 0.095, 1.098 \}$. We model the dielectric constant of
Au by a Drude model $\varepsilon(i\xi) = 1 + \omega^2_p / \xi
(\xi + \gamma)$, where $\omega_p = 1.367 \times 10^{14}$~Hz and
$\gamma = 5.320 \times 10^{13}$~Hz~\cite{Bergstrom97}. The Casimir
force $\vec{F}$ is computed as an integral 
$\vec{F}\sim\int_0^\infty d\xi \oiint \langle \vec{T} \rangle
d\vec{A}$~\cite{Rodriguez07:PRA}, where the mean stress tensor
$\langle \vec{T} \rangle$ is computed at each position and frequency
from the Green's function as described in
\citeasnoun{Rodriguez07:PRA}.  The Casimir torque $\boldsymbol{\tau}$,
was computed in the same framework, with a minor modification,
$\boldsymbol{\tau} \sim \int_0^\infty d\xi \oiint \vec{r} \times
(\langle \vec{T} \rangle d\vec{A})$, also proposed in our earlier
work~\cite{Rodriguez07:PRA}.

We are hopeful that these phenomena will be amenable to experiment.
There are some experimental advantages over planar geometries: the
concentric configuration is a stable equilibrium, static charges on
the outer metallic cylinder are screened in the interior, and the
equilibrium is easily distinguished from electrostatic effects (which
cannot produce stability~\cite{Dalvit04}).  (If we inverted the
geometry, to have the SiO$_2$ on the outside and the metal on the
inside, then the larger static $\varepsilon$ of the fluid compared to
the SiO$_2$ would lead to a classical stable equilibrium for a charged
inner cylinder~\cite{Chiao94, Milonni95, Dalvit04}.)  We expect that
similar stability will be obtained for real metals, as in our Au
calculation in \figref{1drep}, and there are several materials that
exhibit repulsive-attractive transitions (unlike SiO$_2$--ethanol--Au)
that should display the orientation transition. We have also uncovered an intriguing
question to explore in future work: in the PFA heuristic, it appears
that the critical lengthscale at which the orientation transition occurs is \emph{smaller}
than the corresponding positional transition lengthscale (changing
from stable to unstable suspension), whereas the exact calculations
give the opposite result.

This work was supported in part by U.~S. Dept. of Energy Grant
\#DE--FG02-97ER25308, the NSF MRSEC program under Grant
\#DMR-0213282, and by the MIT Ferry Fund.


\end{document}